\documentclass[aps,prl,twocolumn,superscriptaddress,showpacs]{revtex4}

\usepackage{graphicx}
\usepackage{dcolumn}
\usepackage{bm}
\usepackage{color}

\begin{document}

\title{Bifurcation in epigenetics: implications in development, proliferation and diseases}

\author{Daniel Jost}

\affiliation{Laboratoire de Physique, \'Ecole Normale Sup\'erieure de Lyon, CNRS UMR 5672, Lyon, France}

\begin{abstract}
Cells often exhibit different and stable phenotypes from the same DNA sequence. Robustness and plasticity of such cellular states are controlled by diverse transcriptional and epigenetic mechanisms, among them the modification of biochemical marks on chromatin.
Here, we develop a stochastic model that describes the dynamics of epigenetic marks along a given DNA region. Through mathematical analysis, we show the emergence of bistable and persistent epigenetic states from the cooperative recruitment of modifying enzymes. We also find that the dynamical system exhibits a critical point and displays, in presence of asymmetries in recruitment, a bifurcation diagram with hysteresis. These results have deep implications for our understanding of epigenetic regulation. In particular, our study allows to reconcile within the same formalism the robust maintenance of epigenetic identity observed in differentiated cells, the epigenetic plasticity of pluripotent cells during differentiation and the effects of epigenetic misregulation in diseases.  Moreover, it suggests a possible mechanism for developmental transitions where the system is shifted close to the critical point to benefit from high susceptibility to developmental cues. 
\end{abstract}


\maketitle

Cellular differentiation occurs during the development of multicellular organisms and leads to the formation of many different tissues where gene expression is modulated without modification of the genetic information \cite{AllisBook}. These modulations are in part encoded by biochemical tags, called epigenetic marks, that are set down at the chromatin level directly on DNA or on histone tails. These marks are directly or indirectly involved in the local organization and structure of the chromatin fiber, and therefore may modulate the accessibility of DNA to transcription factors or enzymatic complexes, playing a fundamental role in the transcriptional regulation of gene expression.

Each tissue is characterized by a distinct epigenetic pattern \cite{Doi2009} that is mainly shaped during cellular differentiation by developmental signals driven mainly by transcription factors. For a differentiated cell, these specific signals disappear and the global epigenetic state of a cell is robustly maintained throughtout the cell life and in its daughter cells. This maintenance, despite the fast turnover rate of epigenetics marks \cite{Deal2010} or the dilution of epigenetic information during cell division \cite{Margueron2010}, implies the existence of mechanisms to avoid the rapid loss of epigenetic information.  

At the gene level, there are active or inactive epigenetic marks that influence the transcriptional activity of the gene \cite{Jenuwein2001}. A coherent activity needs that the gene promoter is covered by a majority of active or inactive marks.
Recently, many efforts both experimentally \cite{Kelemen2010,Angel2011,Hathaway2012} and theoretically \cite{Dodd2007,DavidRus2009,Mukhopadhyay2010,Micheelsen2010,Dodd2011,Becksei2011,Sneppen2012,Satake2012,Binder2013, Dayarian2013} have been dedicated to study the mechanisms responsible for the spreading and maintenance of an epigenetic state at the gene level for different biological organisms and contexts. In particular, it has been shown for many examples, like for the locus {\it MAT} in fission yeast \cite{Dodd2007, Micheelsen2010} or for the vernalization of the floral repressor {\it FLC} in {\it Arabidopsis} \cite{Angel2011,Satake2012}, that long-range or cooperative interactions between epigenetic marks are necessary for the emergence of stable coherent states. Of particular interest is also the work of the Sengupta's group showing that differences between activating and silencing rates might be responsible for hysteresis in epigenetic silencing by the SIR system in budding yeast \cite{Mukhopadhyay2010,Dayarian2013}.

In this Rapid Communication, we develop a general formalism that describes the dynamics of epigenetic marks and that accounts, at the same time, for the plasticity and for the robustness of the epigenome. Using statistical and non-linear physics methods, we characterize the emergence of coherent epigenetic states from the recruitment of modifying enzymes, and we study the bifurcations occuring in the system when altering the recruitment intensities. Finally, we discuss in details the implications of our findings in concrete biological contexts during development, differentiation or disease. In particular, we propose that the experimentally-observed regulation of the cell cycle during development might have a strong impact on the efficiency of epigenetic switches.

\paragraph{Model.} Inspired by the model introduced by Dodd et al for the mating-type switch of the fission yeast ({\it S. pombe}) \cite{Dodd2007,DavidRus2009,Micheelsen2010}, we consider a DNA region, consisting of $n$ nucleosomes, that is located between two boundaries that epigenetically isolate the region from neighboring DNA \cite{Gaszner2006,Dodd2011}. We assume that the epigenetic state of each nucleosome can fluctuate between three different states: unmarked (U), active (A) and inactive (I). Active marks would be for example associated with acetylation of the lysine 9 of histone 3 (H3K9), and inactive marks with tri-methylation of H3K9 (HP1-type chromatin) or with the methylation of H3K27 (Polycomb-type chromatin) \cite{Jenuwein2001}. However, the actual epigenetic marks associated with each state are not important for our purpose and we only need two well separated kinds of marks that can be exchanged by passing through an intermediate state to each other by passing through an unmarked state ($I\rightleftharpoons U \rightleftharpoons A$).

We assume that the nucleosomal state can be modified by two mechanisms: i) Recruitment of modifying enzymes (for example, histone demethyl- or methyltransferases, and histone deacetyl- or acetyltransferases) by surrounding active or inactive nucleosomes that occurs at a rate $\epsilon_X \rho_X$ where $\rho_X$ is the local density of the modified state X (A or I) that, by considering a spatial mean-field approximation, we identified to the {\em global} density $n_X/n$ with $n_X$ the corresponding number of nucleosomes. ii) Random transitions between states that occur at a rate $k_0$ and that represents recruitment-independent enzymatic activity, nucleosome turnover or dilution due to replication. Note that to facilitate the analysis we lumped into $k_0$ processes (turnover vs replication) with presumably different time-scales and statistical properties. However previous works on epigenetic modeling \cite{Dodd2007,DavidRus2009} suggest that the main conclusions of our study should not depend on that approximation. 
The corresponding system of biochemical reactions is composed of four possible state transitions given below with their respective propensities:
\begin{eqnarray}
U\rightarrow A & & \quad r_{u,a}\equiv (k_0+\epsilon_A \rho_A)(n-n_A-n_I),\label{eq:e1}\\
U\rightarrow I & & \quad r_{u,i} \equiv (k_0+\epsilon_I \rho_I)(n-n_A-n_I),\label{eq:e1b}\\
A\rightarrow U & &\quad r_{a,u}\equiv (k_0+\epsilon_I \rho_I)n_A, \label{eq:e2b}\\
I\rightarrow U & &\quad r_{i,u} \equiv (k_0+\epsilon_A \rho_A)n_I,\label{eq:e2}
\end{eqnarray}
where, for simplicity, we assumed that the rates of random transitions are similar for each reaction and that possible discrepancies between $A$ and $I$ occur at the recruitment level.

\paragraph{Analogy with an Ising model and phase transitions.} From a theoretical perspective, we remark that this model is formally very similar to a zero-dimensional 3-states Ising model where nucleosomal states represent spins (for example $I=-1$, $U=0$, $A=+1$), recruitment ($\epsilon_X$) corresponds to coupling between spins ($J$) and random transitions ($k_0$) are associated with thermal fluctuations ($k_B T$). 

This analogy suggests that a good observable for our system will be the magnetization $m=(n_A-n_I)/n$. In biological terms, this magnetization could be interpreted as the relative activity of the DNA region if we consider that, in addition to the favorable effect of active marks, the presence of inactive marks penalizes the activity. In the following, we will consider $m$ as the relevant observable of our system. 

\begin{figure}[t]
\begin{center}
\includegraphics[width=8.5cm]{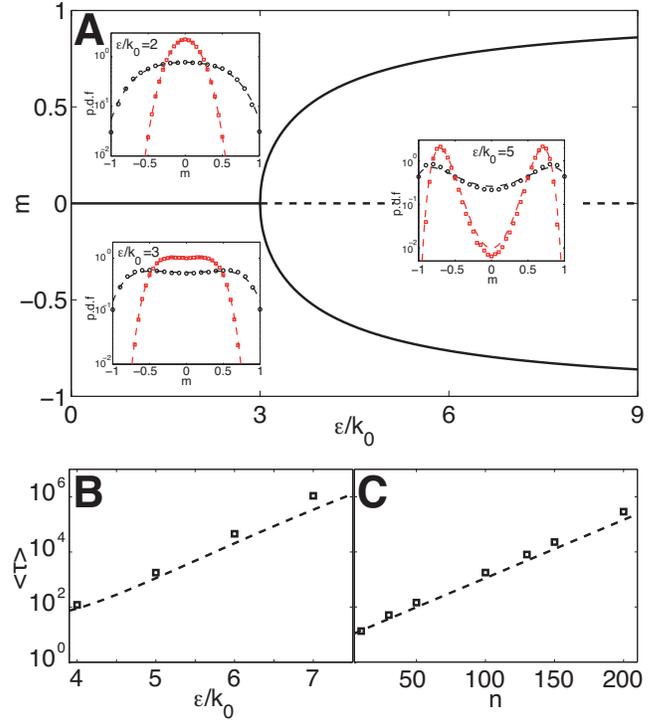}
\end{center}
\caption{(Color online) {\bf Symmetric regime.} (A) Bifurcation diagram for $m$ as a function of $\epsilon/k_0$. Full (dashed) lines represent stable (unstable) fixed points of the dynamical system. (Insets) Probability distribution functions (p.d.f.) of $m$ for $n=10$ (black circles) or $n=100$ (red squares) computed from the Fokker-Planck approximation (dashed lines) or from stochastic simulations (squares). (B,C) Mean first passage time $\langle \tau \rangle$ (in $k_0^{-1}$-unit) to switch from $m_-$ to $m_+$ as a function of $\epsilon$ (B, for $n=100$) or of $n$ (C, for $\epsilon/k_0=5$), computed from Eq.\ref{eq:tau} (dashed lines) or from simulations (squares). Standard errors on the estimation of $\langle \tau \rangle$ are smaller than the symbol size.}\label{fig:sym}
\end{figure}

As it is well known that zero-dimensional Ising models do exhibit phase transitions between an ordered and a disordered phases \cite{Mukamel2005}, we expect our system to display such dramatic changes. To illustrate that we consider the simple mass-action model (equivalent to the mean-field approximation of the Ising model) that captures the mean dynamics of the epigenetic marks in the case of symmetric recruitment ($\epsilon_A=\epsilon_I\equiv \epsilon$):
\begin{eqnarray}
\frac{d\rho_A}{dt} & = & (k_0+\epsilon \rho_A)(1-\rho_A-\rho_I)-(k_0+\epsilon \rho_I)\rho_A, \label{eq:ma1}\\
\frac{d\rho_I}{dt} & = & (k_0+\epsilon \rho_I)(1-\rho_A-\rho_I)-(k_0+\epsilon \rho_A)\rho_I. \label{eq:ma2}
\end{eqnarray}
At steady-state, the previous dynamical system has at most three relevant fixed points that, in term of the effective magnetization $m$, are given by $m_0=0$ ($\forall \epsilon$) and $m_{\pm}=\pm (k_0/\epsilon)\sqrt{(\epsilon/k_0+1)(\epsilon/k_0-3)}$ (for $\epsilon>3k_0$). Fig.\ref{fig:sym}A shows the classical supercritical pitchfork bifurcation occuring at the critical point $\epsilon_c=3k_0$. 
For weak recruitment ($\epsilon < \epsilon_c$), $m_0$ is the only stable fixed point and the activity of the DNA region is not clearly defined.  At the critical point, $m_0$ turns unstable and stable coherent activities ($m_{\pm}$ for active or inactive) are only observed for strong recruitments ($\epsilon > \epsilon_c$). 

\paragraph{Distribution and stability of epigenetic states.} To go beyond the mean-field approximation, we aim at the full distribution of probability for $m$. For simplicity, we assume symmetric coupling in the following. The general situation will be discussed in the next section.
The starting point is to write the master-equation related to the set of biochemical reactions given in Eqs.\ref{eq:e1}-\ref{eq:e2}, but for the new variables $m$ and $s=(n_A+n_I)/n$ where $s$ represents the density of marked nucleosomes.
Then, assuming that $s$ is always large and that its dynamics is fast compared to the one of $m$ allows to perform a time-scale separation and leads, in the limit of large $n$, to the Fokker-Planck equation for the probability $P(m)$ (see \cite{supmat} for details)
\begin{eqnarray}
\frac{\partial P}{\partial t} & = & -\frac{\partial}{\partial m}\left([w_+(m)-w_-(m)]P \right. \nonumber \\
 & & \left. -(1/2n)\frac{\partial}{\partial m}\left([w_+(m)+w_-(m)]P\right)\right). \label{eq:pinf}
\end{eqnarray}
with $w_+=(r_{u,a}+r_{i,u})/n$ and $w_-=(r_{u,i}+r_{a,u})/n$, the propensities to increase (using reactions \ref{eq:e1} and \ref{eq:e2}) or decrease (\ref{eq:e1b} and \ref{eq:e2b}) $m$ by 1.
Eq. \ref{eq:pinf} is a classical Fokker-Planck equation for the heterogeneous diffusion of a particle within an unidimensional potential \cite{RiskenBook}. At steady-state, the probability distribution is then given by
\begin{equation}\label{eq:pinf}
P_{\infty}(m)=\frac{1}{Z}\frac{\exp\left[2n\int_{-\infty}^m dm'\left(\frac{w_+(m')-w_-(m')}{w_+(m')+w_-(m')}\right)\right]}{w_+(m)+w_-(m)}
\end{equation}
with $Z$ a normalization factor. Insets of Fig.\ref{fig:sym}A shows the very good agreement between the approximated solutions given by Eq.\ref{eq:pinf} and the distributions computed from exact stochastic simulations of the full system (\ref{eq:e1}-\ref{eq:e2}) using the Gillespie algorithm \cite{Gillespie1977}. While below the critical point, we observe a unimodal distribution centered around $m_0$, for high coupling ($\epsilon>\epsilon_c$), the distribution is bimodal and peaked around the coherent epigenetic states $m_{\pm}$, the width of each peak being proportional to $[n(\epsilon-\epsilon_c)/k_0]^{-1/2}$ \cite{supmat}. At the critical point, we observe a nearly flat distribution characteristics of phase transitions. 

In presence of bimodality, we quantify the stability of a coherent epigenetic state by computing the mean first passage time $\langle \tau \rangle$ to switch from $m_-$ to $m_+$. Using the Fokker-Planck formalism introduced above, we show that \cite{supmat}
\begin{equation}
\langle \tau \rangle \approx  \frac{18 \pi}{(\epsilon-\epsilon_c)\sqrt{3(\epsilon/k_0+3)}}\exp[ V(0)-V(m_-)] \label{eq:tau}
\end{equation} 
with $V(m)=-\log P_{\infty}$ the effective potential corresponding to the steady-state "epigenetic landscape". As already pointed out by Micheelsen and al. \cite{Micheelsen2010}, we find that, similar to transition state theory \cite{LaidlerBook}, $\langle \tau \rangle$ mainly depends on the "energy barrier" between the starting state ($m=m_-$) and the transition state ($m=0$). As the coupling increases, the barrier is higher and the epigenetic state is more stable (see Fig. \ref{fig:sym} B). We also remark that $\langle \tau \rangle$ scales exponentially with the size $n$ of the system (see Fig. \ref{fig:sym} C). Such very long relaxation time are typical to system with long-range interactions \cite{Mukamel2005, Bouchet2010} and means that the larger the system, the more stable the epigenetic state.

\paragraph{Asymmetry and cusp catastrophe.} 

\begin{figure}[t]
\begin{center}
\includegraphics[width=8.5cm]{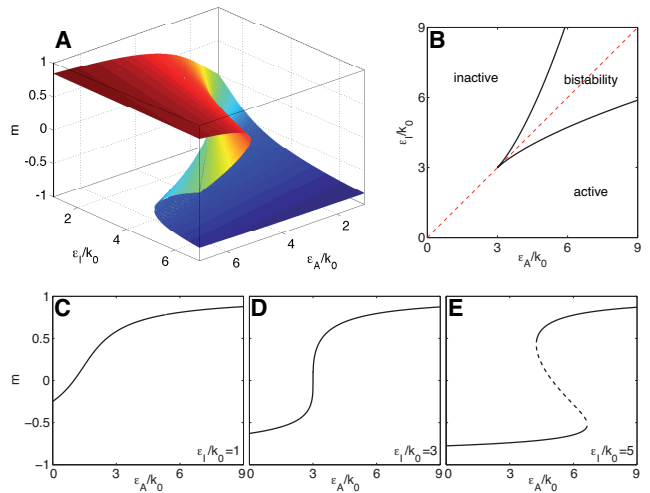}
\end{center}
\caption{(Color online){\bf Asymmetric regime.} (A) Cusp catastrophe surface representing the fixed points of the dynamical system as a function of $\epsilon_A/k_0$ and $\epsilon_I/k_0$. (B) Stability diagram and boundaries between the mono- and bistable regions. (C,D,E) Bifurcation diagrams for $m$ as a function of $\epsilon_A/k_0$ for fixed values of $\epsilon_I/k_0$(=1, C; =3, D; =5, E). Legend as in Fig.\ref{fig:sym}A.}\label{fig:asym}
\end{figure}

In this section, we consider the general situation where recruitments of enzymes by active or inactive marks are different. In the next, we will focus on the study of the generalization of the dynamical system Eq.(\ref{eq:ma1}-\ref{eq:ma2}) that captures the main characteristics of asymmetric recruitment.

At steady-state, the system has at most three fixed points and the bifurcation diagram shapes as a cusp catastrophe surface (Fig. \ref{fig:asym}A), characteristics of dynamical systems with asymmetry \cite{StrogatzBook} and which is also observed for example in the epigenetic SIR system in budding yeast \cite{Mukhopadhyay2010,Dayarian2013} or in insect outbreaks \cite{Ludwig1978}. For every pairs of parameters $(\epsilon_A,\epsilon_I)$, the dynamical system is either monostable or bistable (with an unstable fixed point). Using equality conditions on the nullclines and their first derivatives \cite{supmat}, we find an exact parametric expression for the boundary between the mono- and bistable regions (Fig.\ref{fig:asym}B). Depending on the relative asymmetry between $\epsilon_A$ and $\epsilon_I$, the single (stable) fixed point of the monostable region corresponds to an active or inactive epigenetic state. Bistability, with the coexistence of an active and of an inactive coherent activity, is observed only for strong recruitments ($\epsilon_A,\epsilon_I>\epsilon_c$) and small asymmetry. 

Figures \ref{fig:asym} C-E show the bifurcation diagram of the system when increasing $\epsilon_A/k_0$ for fixed values of $\epsilon_I/k_0$. If $\epsilon_I<\epsilon_c$, the system stays in the monostable region and the unique fixed point goes continously from an almost incoherent state ($m\sim 0$) to a coherent active state (Fig. \ref{fig:asym}C). For stronger recruitment ($\epsilon_I>\epsilon_c$), the system crosses the bistable region making a typical hysteresis curve with two saddle-node bifurcations (Fig.\ref{fig:asym}E). For example, starting from an inactive state, as we increase the recruitment of active marks, the system stays in the inactive state (even for $\epsilon_A>\epsilon_I$) until it switches abruptly to an active state. When crossing the cusp point (Fig.\ref{fig:asym}D), the system become ultra-sensitive and weak asymmetries lead to important changes in the epigenetic state.

\paragraph{Implications in development, proliferation and diseases.}

\begin{figure}[t]
\begin{center}
\includegraphics[width=8.5cm]{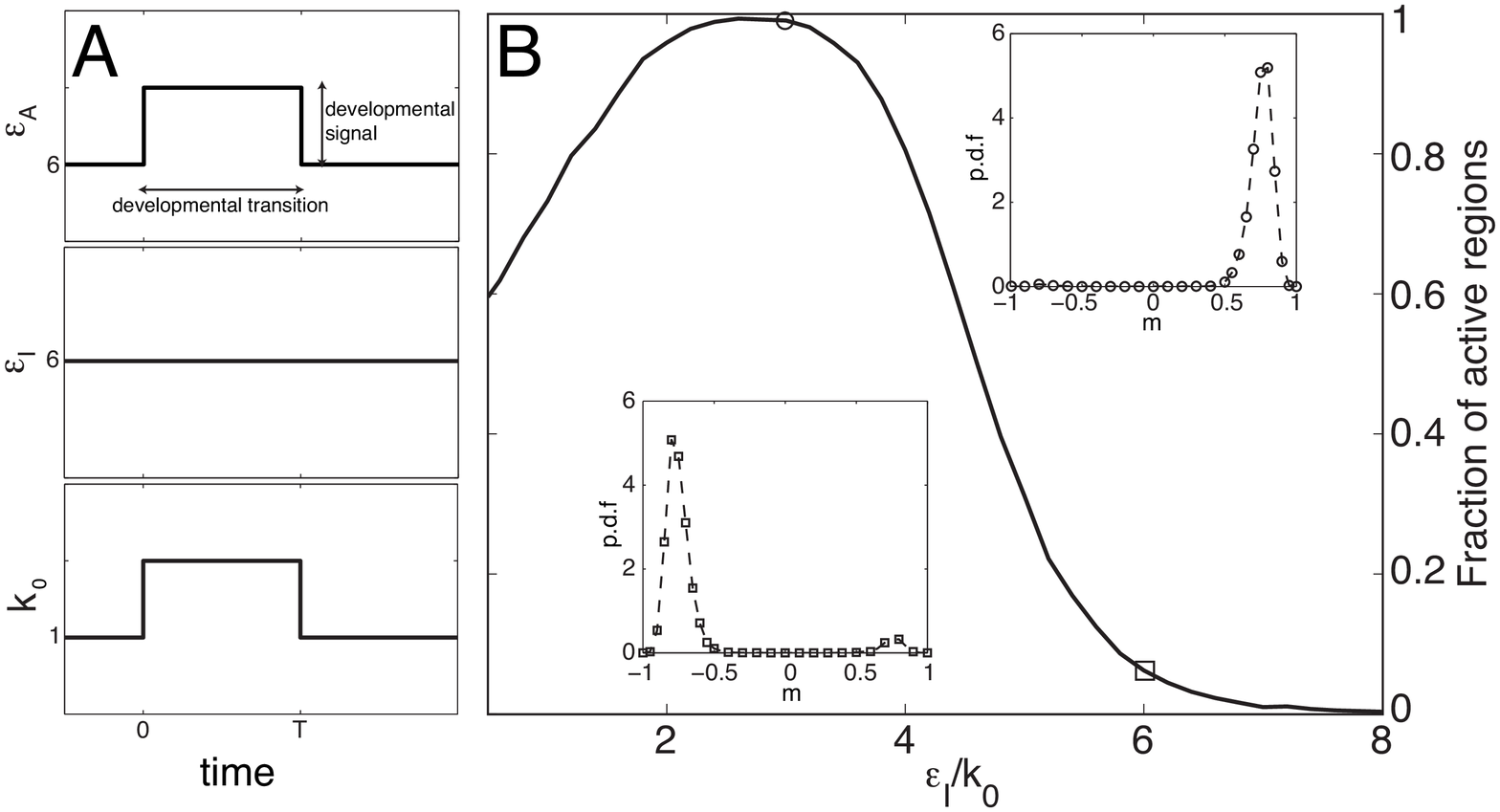}
\end{center}
\caption{(Color online) {\bf Epigenetics and criticality.} (A) Proposed strategy for guided epigenetic switching at developmental transitions when temporary asymetric signal ($\epsilon_A>\epsilon_I=6$) and modification of $k_0$ are applied during a finite time period $T$. (B) Probability to be active ($m>0$) after a time $t=100$ starting from a silenced state ($m=m_-$) as a function of the effective recruitment strength $\epsilon_I/k_0$ for $\epsilon_A=8$ and $T=5$. (Insets) Corresponding probability distribution function of $m$ at $\epsilon_I/k_0=3$ (circles) or $6$ (squares).}\label{fig:dev}
\end{figure}

Recent quantitative experiments in mouse \cite{Hathaway2012} have shown that bistable activities take place in an epigenetically-controled system and that modifications in the epigenome can be stably imprinted by transcriptional factors. In particular, this study suggests that, at least for some chromatin regions, symmetric or weakly asymmetric recruitment may correspond to normal biological situations, and that temporary strong asymetries in the recruitment of modifying enzymes may lead to long-term modifications of the current epigenetic state.

In vivo, differentiated cells exhibit a robust phenotype within the population and in time. The analysis of the symmetric regime of our formalism (Fig. \ref{fig:sym}) suggests that such robustness needs strong recruitments ($\epsilon>\epsilon_c$) and large cooperative units ($n\gg 1$) in order to stabilize coherent epigenetic states (active or inactive) and to avoid spurious switches between coherent states. 
Moreover, Fig. \ref{fig:asym} suggests that epigenetic states of differentiated cells are also stable against small fluctuations of the recruitment couplings. Such fluctuations induce asymmetries in the system that do not impact significantly on the epigenetic state as long as $\epsilon_A$ and $\epsilon_I$ remains in the bistability region. This property is crucial for the maintenance of a robust phenotype in weakly fluctuating environments.

However, when modifications of the environment are important, it would be beneficial to adapt to the current environment by modifying the epigenetic state, like for example for plants at seasonal transitions \cite{Horvath2009}. Misadaptation to the current environment may lead to stress signalings that might result in asymetric recruitment forcing the current epigenetic state towards an adapted state, like in the cold-induced silencing of the {\it FLC} locus in plant vernalization \cite{Angel2011}.
Fig.\ref{fig:asym} E suggests that such asymmetric signals have to be strong enough to allow an epigenetic switch. Once the switch is performed, the hysteretic shape of the bifurcation insures that the epigenome would remain stably in its new state. This property allows adaptation but only if needed, i.e. only when the organism is strongly misadapted to the current environment. 

Many diseases have been related to epigenetic perturbations, from neurologic disorder to cancer \cite{Portela2010}. Within our formalism, these perturbations could be interpreted by anomalous values for the recruitment ($\epsilon/k_0$) that shift the steady-state of the system below or close to the critical point, and that make the epigenetic state uncoherent or unstable and very sensitive to external noise. For example, cancer is often associated with an increase in the frequency of replication during tumorigenesis \cite{Malumbres2009}. In our model, this means an increase of the random transition rate $k_0$ due to replication. This may modify the position of the critical point and therefore may lead to epigenetic instability and misregulation of some tumor suppressor proteins for example. 

During development, cellular differentiation occurs in successive stages. Cells pass through series of developmental transitions where epigenetic states are modified by developmental cues that presumably force locally the desired state. Previously, we saw that, for differentiated cells, the hysteretic shape of the bifurcation may be valuable for the buffering of environmental fluctuations or for adaptation. Within the context of development, this could represent a hindrance at developmental transitions when the epigenetic state has to efficiently switch in a short time window. 
Our formalism suggests that a possible stategy to overcome this apparent issue would be to shift temporarily the system at or close to the critical point during developmental transition. Indeed, Fig.\ref{fig:asym}D shows that at the cusp point, the epigenetic state is very sensitive to weak asymmetries. Going back to the original analogy between our model and an Ising model with phase-transition, this is related to the concept of susceptibility that is maximal at the critical point. 
In many organisms, experimental evidence suggest that the cell cycle length is regulated during development alternating between short cycles during developmental transitions and longer cycles between the transitions and in the differentiated state \cite{Budirahardja2009}. 
In our model this means that at developmental transitions, $k_0$ may be increased and therefore that the effective recruitment strength ($\epsilon/k_0$) may be reduced and may become closer to the critical value. Under this hypothesis, we test the ability to switch between two coherent states when applying a weak asymmetric signal (the developmental signal) during a finite time period (the developmental transition), by running Gillespie simulations. As expected, Fig.\ref{fig:dev} shows that the switching efficiency is optimal when $\epsilon_I/k_0\sim  3$ during the transition. Compared to the situation where $k_0$ is not changed during the transition and where only few cells have switched their epigenetic state, going close to the critical point leads to stable switch for almost all the cells.
Provided the necessary experimental verifications of the proposed strategy, our results strongly suggest that it could be advantageous, during developmental transitions, of being close to criticality to benefit from the high sensitivity to external stimuli, as already observed in various other biological systems \cite{Mora2011}.

\begin{acknowledgements}
I thank C\'edric Vaillant for fruitful discussions.
\end{acknowledgements}

\end{document}